\documentclass[12pt]{article}
\usepackage{amssymb}
\usepackage{epsfig}
\usepackage{float}
\newcommand{\be}{\begin{equation}}
\newcommand{\ee}{\end{equation}}
\newcommand{\bea}{\begin{eqnarray}}
\newcommand{\eea}{\end{eqnarray}}

\def\gtap{\mathrel{ \rlap{\raise 0.511ex \hbox{$>$}}{\lower 0.511ex
   \hbox{$\sim$}}}} 
\def\ltap{\mathrel{ \rlap{\raise 0.511ex
   \hbox{$<$}}{\lower 0.511ex \hbox{$\sim$}}}}
\begin{document}

\begin{flushright}
SISSA 18/2006/EP \\
hep-ph/0603178
\end{flushright}

\begin{center}
{\bf A COMMENT ON THE MEASUREMENT OF NEUTRINO MASSES 
IN $\beta$-DECAY EXPERIMENTS}
\end{center}

\vspace{0.6cm}
\begin{center}
S. M. Bilenky~$^{{\rm a,b}}$, M. D. Mateev~$^{{\rm c}}$ and 
S. T. Petcov~$^{{\rm b,d}}$~
\footnote{Also at: Institute of Nuclear Research and
Nuclear Energy, Bulgarian Academy of Sciences,
1784 Sofia, Bulgaria.}
\end{center}

\begin{center}
{\em  $^{{\rm a}}$Joint Institute
for Nuclear Research, Dubna, R-141980, Russia.} 
\end{center}

\vspace{-0.8cm}
\begin{center}
{\em $^{{\rm b}}$Scuola Internazionale Superiore di Studi Avanzati,
I-34014 Trieste, Italy.}
\end{center}

\vspace{-0.8cm}
\begin{center}
{\em  $^{{\rm c}}$ University of Sofia "St. Kliment Ohridsky", 1164 Sofia, 
Bulgaria.}
\end{center}

\vspace{-0.8cm}
\begin{center}
{\em  $^{{\rm d}}$Istituto Nazionale di Fisica Nucleare,
I-34014 Trieste, Italy.}
\end{center}

\begin{abstract}
We discuss the physics potential of 
future tritium $\beta$-decay experiments
having a sensitivity to a neutrino mass 
$\sim \sqrt{|\Delta m^{2}_{23}|} \sim 5\times 10^{-2}$ eV.
The case of three-neutrino mixing is analised.
A negative result of such an experiment 
would imply that the neutrino mass spectrum 
is of normal hierarchical type.
The interpretation of a positive result 
would depend on the value of the lightest neutrino mass;
if the lightest neutrino mass satisfies
the inequality ${\rm min}(m_j) \ll \sqrt{|\Delta m^{2}_{23}|}$,
it would imply that the neutrino mass
spectrum is of the inverted hierarchical type.

\end{abstract}

\section{Introduction}

\hskip 1.0cm The discovery of neutrino oscillations in the 
experiments with solar, atmospheric and reactor neutrinos
\cite{Cl,Gallex,Sage,SKsol,SK,SNO,Kamland,K2K}
is the first particle physics evidence
for existence of a new 
beyond the Standard Model physics.
The solar, atmospheric, reactor and K2K neutrino data 
imply the presence of 3-$\nu$ mixing
in the weak charged lepton current:
\be
\nu_{lL}(x) = \sum_{i=1}^{3} U_{li} \nu_{iL}(x),~~l=e,\mu,\tau \;.
\label{1} 
\ee
%
\noindent Here $\nu_{lL}(x)$ is the mixed flavor neutrino field,
$U$ is the 3$\times$3 unitary PMNS \cite{BP,MNS} mixing matrix 
and $\nu_{iL}(x)$ is the field of neutrino with mass $m_{i}$. 
All currently existing $\nu$-oscillation data, except
the data of the LSND experiment  
\footnote{In the
LSND experiment indications 
for oscillations
$\bar \nu_{\mu}\to\bar \nu_{e}$  
with $(\Delta m^{2})_{\rm{LSND}}\simeq 
1~\rm{eV}^{2}$ were obtained. 
The  LSND results are being tested 
in the MiniBooNE experiment
\cite{MiniB}.}
\cite{LSND},
can be described perfectly well 
assuming  3-$\nu$ mixing in vacuum
and we will consider this possibility in what follows.
 The minimal 4-$\nu$ mixing scheme 
which could incorporate the LSND indications for 
$\nu$-oscillations is strongly disfavored by the data. 
The $\nu$-oscillation explanation of the LSND results
is possible assuming 5-$\nu$ mixing \cite{JConrad}.

The experimental study of neutrino oscillations allowed to 
determine the values of the two neutrino mass-squared 
differences $\Delta m^{2}_{12}$ and   
$|\Delta m^{2}_{23}|$ ($\Delta m^{2}_{ik}= m^{2}_{k} - m^{2}_{i} $) 
and to obtain information on the three 
neutrino mixing angles $\theta_{12}$, 
$\theta_{23}$ and $\theta_{13}$, which characterise
the oscillations in the case of 
3-neutrino mixing. From analysis of the Super-Kamiokande 
atmospheric neutrino data, the following 
best fit values and 
95\% CL allowed ranges of values
of the parameters $|\Delta m_{23}^{2}|$ and 
$\sin^{2}2\theta_{23}$
were obtained \cite{SK,TSchw05}:
 \bea 
|\Delta m^{2}_{23}| \cong 2.4\cdot 10^{-3}\rm{eV}^{2},
~~ \sin^{2}2 \theta_{23} = 1.0\;, \\ [0.25cm]
1.7\cdot 10^{-3}\leq |\Delta m^{2}_{23}| \leq 2.9\cdot 10^{-3}\rm{eV}^{2},
~~ \sin^{2}2 \theta_{23} > 0.90.
\label{2} 
\eea 
%
\noindent The combined analysis of the 
solar neutrino and KamLAND
data allowed to determine $\Delta m^{2}_{12}$
and $\sin^{2} \theta_{12}$. 
For the best fit values and 95\% 
allowed ranges it was found \cite{SNO,Kamland,BCGPRKL2}
\bea 
\Delta m^{2}_{12} = 8.1\cdot 10^{-5}\rm{eV}^{2},
~~ \sin^{2}\theta_{12} = 0.31\;, \\ [0.25cm]
7.3\cdot 10^{-5}\leq \Delta m^{2}_{12} \leq 9.0\cdot 10^{-5}\rm{eV}^{2},
~~ 0.26 \leq \sin^{2} \theta_{12}\leq 0.37.
\label{3} 
\eea
%
\noindent Only an upper bound 
on the mixing angle $\theta_{13}$ 
has been obtained so far.
A combined analysis of the data of the CHOOZ experiment
with reactor neutrinos \cite{Chooz} 
and the data from solar neutrino and KamLAND experiments
leads to the following limit  
\cite{BCGPRKL2}   
\be 
\sin^{2}\theta_{13} < 0.027~(0.044)\;,~~~~~~~95\%~(99.73\%)~{\rm C.L.} 
\label{4} 
\ee
%
\indent   The existing atmospheric neutrino data does not 
allow to determine the ${\rm sgn}(\Delta m_{23}^{2})$.
As a consequence, two types of neutrino mass 
spectrum in the case of three-neutrino mixing 
are possible (see, e.g., \cite{STPNu04}):
\begin{itemize}
\item with normal mass ordering (or hierarchy) 
corresponding to $\Delta m_{23}^{2} > 0$,
\be
m_{1}< m_{2} <  m_{3},~\Delta m^{2}_{12} \ll \Delta m^{2}_{23}\;,
\label{5}
\ee
\item with inverted mass ordering (or hierarchy),
associated with $\Delta m_{23}^{2} < 0$,
\be
m_{3}<  m_{1}    <  m_{2},~ \Delta m^{2}_{12}  \ll
 |\Delta m^{2}_{23}| \cong |\Delta m^{2}_{13}|\;. 
\label{6}
\ee
\end{itemize}

  The absolute values of neutrino masses $m_{j}$ 
are unknown at present. In particular, 
for each of the two types of 
the neutrino mass ordering 
there can be strong hierarchy
between the masses, or the splitting between 
the masses can be much smaller than 
the absolute values of the masses.
Correspondingly, 
depending on the sign of 
$\Delta m_{23}^{2}$
and the value of the lightest neutrino mass,  
${\rm min}(m_j)$, 
the $\nu$-mass  spectrum can be (see, e.g., \cite{STPNu04,Altarelli}):
\begin{itemize}
\item
{\it Normal Hierarchical (NH)} 
\be
m_1 \ll m_2 \ll m_3\;:~~
m_2 \cong \sqrt{\Delta m_{12}^{2}} \sim 9\cdot 10^{-3}~{\rm eV}\;,~
m_3 \cong \sqrt{\Delta m_{23}^{2}}\sim 4.9\cdot 10^{-2}~{\rm eV}\;; 
\label{NH}
\ee
\item
{\it Inverted Hierarchical (IH)} 
\be
m_3 \ll m_1 < m_2\;:~~
m_{1,2} \cong \sqrt{|\Delta m_{23}^{2}|} \sim 4.9\cdot 10^{-2}~{\rm eV}; 
\label{IH}
\ee
\item
{\it Quasi-Degenerate (QD)}: 
\be
m_1 \cong m_2 \cong m_3 \cong m_0,~~
m_j^2 \gg |\Delta m_{23}^{2}|\;:~~m_0 \gtap 0.10~{\rm eV}\;. 
\label{QD}
\ee
\end{itemize}

   The measurement of the absolute 
values of neutrino masses and the 
determination of the type of 
neutrino mass spectrum is one of the 
highest priority and most difficult 
problems in neutrino physics  (see, e.g., \cite{STPNu04}). 
The solution of this problem is of fundamental
importance 
for the progress in understanding the origin of 
neutrino masses and mixing.

Information about the absolute values of neutrino 
masses can be obtained from:\\
a) precision measurements 
of the $\beta$-spectrum in the end-point region \cite{Fermi,Perrin};\\
b) investigation of neutrinoless double $\beta$-decay, if the 
neutrinos with definite mass are Majorana particles 
\cite{BPP1,STPFocusNu04};\\
c) measurement of power spectrum of the large 
scale distribution of galaxies (see, e.g., \cite{WMAPnu}).

  From the data of the Mainz \cite{Mainz} and 
Troitsk \cite{Troitsk} tritium
$\beta$-decay experiments the
following upper bound 
on the measurable neutrino mass
was obtained (95\% C.L.):
\be 
m_{\beta} \cong m_0 < 2.3\,\rm{eV}\;. 
\label{10} 
\ee 
%
In the future KATRIN experiment \cite{Katrin} 
a sensitivity to 
\be 
m_{\beta} \cong m_0 \cong 0.2 \,\rm{eV}\;, 
\label{11} 
\ee 
%
is planned to be achieved
\footnote{For an alternative method of direct 
neutrino mass measurement, based on a calorimetric study of 
$\beta$-decay of $^{187}$Re, see ref. \cite{Renium}.}.

  Using the Cosmic Microwave Background (CMB) data of the 
WMAP experiment, combined with data from large
scale structure surveys (2dFGRS, SDSS), 
for the sum of neutrino masses upper bounds in the range 
\be 
\sum_{i} m_{i} \leq (0.4 - 1.7)\,\rm{eV}\;, 
\label{12} 
\ee 
%
were found (see, e.g., \cite{Teg2}).
Data on weak lensing of galaxies 
by large scale structure, 
combined with data from the WMAP and PLANCK
experiments, may allow $\sum{} m_{i}$ to be determined 
with an uncertainty of 
$\delta \sim 0.04$~eV \cite{Hann06}.

 The sign of $\Delta m^2_{23}$, which drives the
dominant atmospheric neutrino oscillations, can be 
determined by studying 
oscillations of neutrinos and
antineutrinos, say, 
$\nu_{\mu} \rightarrow \nu_e$
and $\bar{\nu}_{\mu} \rightarrow \bar{\nu}_e$,
in which matter effects are sufficiently large.
This can be done, e.g., in long-baseline 
$\nu$-oscillation experiments 
(see, e.g.,~\cite{AMMS99}).
Information about ${\rm sgn}(\Delta m^2_{23})$
can be obtained also in atmospheric neutrino 
experiments by studying 
the oscillations of the
atmospheric $\nu_{\mu}$ and $\bar{\nu}_{\mu}$ which
traverse the Earth \cite{JBSP203}.

  In the present article we investigate 
the physics potential 
of $\beta$-decay experiments having a sensitivity 
which permits to probe the  neutrino mass
range corresponding to $\sqrt{|\Delta m^{2}_{23}|}
 \cong 5\cdot 10^{-2}~\rm{eV}^{2}$.
At present such a sensitivity 
does not seem reachable in any realistic experiment.
However, this situation may change in the future.
In our analysis we
take into account the existing and prospective 
neutrino oscillation data on the neutrino mass 
squared differences and neutrino mixing angles.  
We consider different possible 
types of neutrino mass spectrum as well 
\footnote{For recent related analyses see, e.g., ref. \cite{3HLisi04}.
}.

\section{On the Measurement of Neutrino Mass in $\beta$-Decay Experiments}

\hskip 1.0cm The measurement of 
$\beta$-spectrum in the end-point region 
in tritium $\beta$-decay,
\be
 ^{3}\rm{H}\to ^{3}\rm{He} + e^- + \bar \nu_{e}\;, 
\label{13} 
\ee 
%
is the classical method of direct 
determination of neutrino mass, 
which originated from
the pioneering articles of Fermi and Perrin
on $\beta$-decay \cite{Fermi,Perrin}.
This decay has many  advantages (see, e.g., \cite{Mainz}). 
First of all, it is a super-allowed decay. 
Thus, the nuclear matrix element
is a constant and the electron spectrum 
is determined by the relevant
phase space factor only. Other  advantages of this decay 
are the relatively small
energy release  ($E_{0}\simeq 18.574.3\pm 1.7 $ eV)  
and convenient half-life (12.3 years).

   Taking into account the neutrino mixing,
for the effective Hamiltonian of the process (\ref{13}) we have
\be
\mathcal{H}_{I}^{\mathrm{CC}}= \frac{G_{F}}{\sqrt{2}}\,2 \sum_{i} U_{ei}~\bar
e_{L}\gamma _{\alpha}\nu_{iL}\,~ J^{\alpha} + \mathrm{h.c.},
\label{14}
\ee
%
where $J^{\alpha}$ is the hadronic charged current.
For the state  vector  
of the final neutrinos and 
electron we obtain from (\ref{14}) 
\be 
|f\rangle=\sum_{i} |\bar\nu_{i}\rangle\,|e^{-}\rangle\, \langle
\bar\nu_{i}\,e^{-}\,^{3}\rm{He}\,| S |\,~^{3}\rm{H}\rangle, 
\label{15} 
\ee
where 
\be \begin{array}{c}
  \langle \bar\nu_{i}\, e^{-}\,^{3}\rm{He}\,| S |\,~^{3}\rm{H} \rangle=  \\
 - i~2\,\frac{G_{F}}{\sqrt{2}}\,N\, U_{ei}~ \bar
u_{L}(p)\,\gamma_{\alpha}\,v_{L}(p_{i})\,\langle ^{3}\rm{He} |\,
J^{\alpha}(0)\,|~^{3}\rm{H}\rangle \,(2\pi)^{4} \delta (P'-P)\;. 
\end{array}
\label{16}
\ee 
%
Here $N$ is the product of standard normalization factors, $p$ is the
momentum of electron, $p_{i}$ is the momentum of antineutrino 
(right-handed neutrino in the Majorana case) 
with mass $m_{i}$, $P$ and  $P'$ are the total initial
and final momenta.

 The final state neutrinos are not detected 
in tritium $\beta$-decay experiments. Taking into
account the orthogonality of the vectors 
$|\bar\nu_{i}\rangle$ and neglecting
the recoil of the $^{3}\rm{He}$ nucleus,
for the electron spectrum we get the incoherent sum 
\be 
\frac{d\,\Gamma}{d\,E_{e}} = \sum_{i}|U_{ei}|^{2}\,~
\frac{d\,\Gamma (m_{i})}{d\, E_{e}}\;, 
\label{17} 
\ee
%
where 
\be
\frac{d\,\Gamma(m_{i})}{d\,E_{e}} =
C\,p_{e}\,(E_{e}+m_{e})\,(E_{0}-E_{e})\,
\sqrt{(E_{0}-E_{e})^{2}-m_{i}^{2}}\,F(E_{e})\,\theta(E_{0}-E_{e}-m_{i})\;.
\label{18} 
\ee 
%
Here $E_{e}\leq E_{0} -m_{i}$ is the kinetic energy of the
electron, $E_{0}$  is the energy released in the decay (\ref{13}),
$p_{e}$ is the electron momentum, $ m_{e}$ is the mass of the electron,
$F(E_{e})$ is the Fermi function which takes into account the Coulomb
interaction of the final state particles, and $C$ is a constant. 
In eq. (\ref{18}) $(E_{0}-E_{e})$ is the neutrino energy and
$p_{i}=\sqrt{(E_{0}-E_{e})^{2}-m_{i}^{2}}$ is 
the momentum of neutrino with mass $m_{i}$.

  Neutrino masses enter into the expression 
for the electron spectrum through
neutrino momenta. It is obvious that the maximal 
distortion of the electron spectrum
can be observed in the region 
$(E_{0}-E_{e}) \sim  m_{i}$, which is less than of
the order of few eV. However, for reasons of luminosity, in
tritium experiments relatively large end-point intervals of the electron
spectrum are measured: in the Mainz experiment \cite{Mainz}
$E_{0}-E_{e}\leq 70$ eV;  
in the future KATRIN experiment \cite{Katrin} the
region $E_{0}-E_{e}\ltap 20$ eV will be explored.

Usually the quantity $m_{\beta}$ defined by 
\be 
m_{\beta}= \sqrt{\sum_{i}|U_{ei}|^{2}~m^2_{i}}\;, 
\label{19} 
\ee 
%
is considered as the neutrino mass related observable in 
$\beta$-decay experiments.
The expression (\ref{19}) is obtained 
(see refs. \cite{Weinheimer1, Vissani1}) 
by developing the neutrino momentum over
$\frac{m_{i}^{2}}{(E_{0}-E_{e})^{2}}$ 
in eq. (\ref{18}). Let us note that in 
the region sensitive to the neutrino 
mass this expansion is not valid, 
while in the region $(E_{0}-E_{e})\gg m_{i}$ 
the effects of neutrino mass can be neglected. 

  For the neutrino mass spectrum 
with normal ordering, the neutrino masses are given 
(in the standardly used convention) by 
\be 
{\rm min}(m_j) = m_1, 
~~m_{2}=\sqrt{m^{2}_{1}+ \Delta m^{2}_{12}},
~~m_{3}= \sqrt{m^{2}_{1}+ \Delta m^{2}_{12} + \Delta m^{2}_{23} }.
\label{20}
\ee
%
In the case of spectrum with inverted 
ordering of neutrino masses we have
\be 
{\rm min}(m_j) = m_3,~~ m_{1}=\sqrt{m^{2}_{3}+ |\Delta m^{2}_{13}|}, 
~~m_{2}= \sqrt{m^{2}_{3}+ |\Delta m^{2}_{13}| + \Delta m^{2}_{12} }\;.
\label{20a}
\ee
%
\indent The neutrino mass-squared 
differences $\Delta m^{2}_{12}$ and 
$|\Delta m^{2}_{23}|$ have been measured in 
the neutrino oscillation experiments. 
The existing  data
allow a determination of 
$\Delta m^{2}_{12}$ and $|\Delta m^{2}_{23}|$ 
at $3\sigma$ with an error of approximately
12\% and 50\%, respectively.
These parameters will be measured
with higher accuracy in the future.
The highest precision in the determination
of $|\Delta m^2_{23}|$
is expected to be achieved 
from the studies
of $\nu_{\mu}$-oscillations in
the T2K (SK) \cite{T2K} experiment:
the 3$\sigma$ uncertainty
in $|\Delta m^2_{23}|$ is estimated to be 
reduced in this experiment to $\sim 6\%$. 

   The unknown parameter in eqs. (\ref{20}) 
and (\ref{20a}) is the lightest neutrino mass $m_{1}$
($m_{3}$). It follows from (\ref{20}) and (\ref{20a}) 
that the minimal value of the heaviest neutrino 
mass in the cases of normal and inverted mass ordering,
$m_{3}$ and $m_{2}$, are given by 
\be
m^{\rm{min}}_{3(2)}=\sqrt{ \Delta m^{2}_{12} + |\Delta m^{2}_{23}|}\simeq
\sqrt{|\Delta m^{2}_{23}|}\;.
\label{21} 
\ee 
%
\indent As we have discussed earlier, 
depending on the value of the lightest neutrino mass,
three types of neutrino mass spectrum 
are usually considered:
NH (normal hierarchical), IH (inverted hierarchical) and
QD (quasi-degenerate), eqs. (\ref{NH}) - (\ref{QD}). 
The QD spectrum is realised if the 
value of the lightest neutrino mass
is relatively large, ${\rm min}(m_j)\gtap 0.1$ eV.
This spectrum requires an approximate
symmetry of the neutrino mass matrix
(see, e.g., \cite{SPAS94,Altarelli}). 
The NH and IH spectra correspond to negligibly small
value of ${\rm min}(m_j)$.
The NH spectrum is 
typically predicted by the GUT models which
unify quarks, charged leptons and neutrinos (see, e.g., 
\cite{Altarelli,GUTM3}). 
The IH spectrum 
can be associated with the existence 
of a broken $L_{e}-L_{\mu}-L_{\tau}$ symmetry 
in the lepton sector \cite{STP82PD}
(see also, e.g., \cite{Altarelli1}).

  The Mainz, Troitsk and KATRIN experiments 
can probe only the quasi-degenerate
neutrino mass spectrum. If in the KATRIN 
experiment, which is under preparation 
at present, a positive effect due to the neutrino 
mass will be observed, we will have
\be 
m_{\beta}\cong  m_{1,2,3} \cong m_0\;.
\label{25}
\ee 
%
In this case no information on the 
${\rm sgn}(\Delta m^{2}_{23})$, i.e., on the
type of ordering of neutrino masses, 
will be obtained.

A negative result of
the KATRIN experiment would imply that 
the neutrino mass spectrum is either NH or IH,
or else is with partial normal or inverted hierarchy
\cite{BPP1}.
It will be crucial in this case to
improve the sensitivity of direct 
neutrino mass measurement experiments
by approximately a factor of 4.
If the sensitivity of 
the $\beta$-decay experiments 
will allow to probe 
values of neutrino masses
$m_{3}^{\rm{min}}(m_{2}^{\rm{min}}) \cong
\sqrt{|\Delta m^{2}_{23}|} \cong
(3.9 - 5.8) \cdot 10^{-2}$ eV,
these experiments will provide fundamental 
information on the absolute scale 
of neutrino masses and on the type of 
neutrino mass spectrum independently 
of the nature of massive neutrinos, which, 
as is well-known, can be Dirac or Majorana particles 
(see, e.g., \cite{BiPet87}).

 Indeed, if the neutrino mass spectrum is 
of the NH type, the contribution of the
heaviest neutrino mass 
$m_{3} \cong \sqrt{|\Delta m^{2}_{23}|}$
to the distortion of the electron spectrum is suppressed
by the factor $|U_{e3}|^{2} = \sin^2\theta_{13} 
< 5\cdot 10^{-2}$ and will be unobservable.
The distortion of the spectrum due to the mass
$m_{2} \cong \sqrt{\Delta m^{2}_{12}} \cong 9\cdot 10^{-3}$ eV,
which is not suppressed by the corresponding 
mixing matrix element, will also be unobservable.
Thus, the electron spectrum that will be observed in the
$\beta$-decay experiments 
in the case of NH neutrino mass spectrum
will effectively correspond to one zero mass neutrino:
\be 
\frac{d\,\Gamma}{d\,E_{e}} \cong 
\frac{d\,\Gamma (m_i=0)}{d\, E_{e}}\;, 
\label{espNH} 
\ee
%
\indent In contrast, if the neutrino mass 
spectrum is of the IH type,
the two heaviest neutrino masses 
$m_{1} \cong m_{2} \cong 
\sqrt{|\Delta m^{2}_{23}|}$ 
will enter into the
expression for the electron spectrum
with the coefficient $1 - |U_{e3}|^{2} \simeq 1$ . 
The spectrum will have the form:
\be 
\frac{d\,\Gamma}{d\,E_{e}} \cong (1 - |U_{e3}|^{2})\,
\frac{d\,\Gamma (m_{1,2})}{d\, E_{e}} +
|U_{e3}|^{2}\,\frac{d\,\Gamma (m_{3}=0)}{d\, E_{e}}\cong 
\frac{d\,\Gamma (\sqrt{|\Delta m^{2}_{23}|})}{d\, E_{e}}\;.
\label{espIH} 
\ee
%

 It follows from the above discussion that the non-observation 
of the effect of neutrino mass in a
$\beta$-decay experiment having a sensitivity to 
$\sqrt{|\Delta m^{2}_{23}|}$
would imply that the neutrino mass spectrum is
of the {\it normal hierarchical} type, i.e., that 
$\Delta m^{2}_{23} > 0$ and $m_1 \ll m_2 \ll m_3$,
independently of whether the massive 
neutrinos are Dirac or Majorana particles.
If the spectrum of neutrino masses is of the 
{\it inverted hierarchical} type,
the effect of neutrino mass must be observed 
in such an experiment.

  The interpretation of a positive result of 
a $\beta$-decay experiment with a sensitivity to
a neutrino mass $\sim \sqrt{|\Delta m^{2}_{23}|}$,
however, will not be unique in what regards the 
${\rm sgn}(\Delta m^{2}_{23})$
and the value of the lightest neutrino mass
(the type of neutrino mass spectrum).
Indeed, in the discussion above of the cases of
NH and IH spectra it was always assumed that 
the lightest neutrino mass is negligible,
i.e., $m_1 \ll \sqrt{\Delta m^{2}_{12}}$ in the NH case and 
$m_3 \ll \sqrt{|\Delta m^{2}_{23}|}$ in the IH one.
However, this may not necessarily be valid. 
In principle, for both normal and inverted neutrino mass ordering,
we can have 
${\rm min}(m_j) \ltap \sqrt{|\Delta m^{2}_{23}|}$,
which corresponds to a spectrum with partial hierarchy \cite{BPP1}.
Thus, the distortion of the electron spectrum 
in the case of positive result of 
a $\beta$-decay experiment under discussion
could be due either to\\
i) spectrum with {\it inverted neutrino mass ordering},
$\Delta m^{2}_{23} < 0$, of two possible types:\\ 
a) {\it inverted hierarchical}, $m_3 \ll m_1 < m_2$, or\\
b) {\it with partial inverted hierarchy}, $m_3 < m_1 < m_2$ \cite{BPP1};\\
or to \\
ii) spectrum with {\it normal neutrino mass ordering},
$\Delta m^{2}_{23} > 0$, but with {\it partial neutrino mass hierarchy},
$m_1 < m_2 < m_3$ \cite{BPP1}.\\

   As an example of the possibility ii) 
consider the following hypothetical spectrum:
$m_{1} = 5.0\cdot 10^{-2}$ eV,
$m_{2} = \sqrt{m_1^2 + \Delta m^2_{12}} \cong 5.1\cdot 10^{-2}$ eV,
$m_{3} = \sqrt{m_1^2 + \Delta m^2_{13}} \cong 6.9\cdot 10^{-2}$ eV,
where we have used the best fit values of 
$\Delta m^2_{12}$ and $\Delta m^2_{13}$ given in 
eqs. (\ref{2}) and (\ref{3}).
The sum of neutrino masses is equal to $\Sigma m_i \cong 0.17$ eV.
Obviously, a $\beta$-decay experiment 
having the precision 
under discussion will not be sensitive 
to the difference between the masses $m_1$ and $m_2$;
and it will not be sensitive to the distortion  
of the electron spectrum due to the mass $m_3$
since the contribution of the latter
is suppressed by the factor $\sin^2\theta_{13}$.
In this case we will have for the electron spectrum,
\be 
\frac{d\,\Gamma}{d\,E_{e}} \cong (1 - |U_{e3}|^{2})\,
\frac{d\,\Gamma (m_{1,2})}{d\, E_{e}} +
|U_{e3}|^{2}\,\frac{d\,\Gamma (m_{3})}{d\, E_{e}}\cong 
\frac{d\,\Gamma (m_{1,2})}{d\, E_{e}}\;,
\label{espPNH} 
\ee
%
which practically coincides with the form of the
electron spectrum predicted in the case
of neutrino mass spectrum of {\it inverted hierarchical} 
type, eq. (\ref{espIH}).

   Let us note that neutrino mass spectrum with partial 
hierarchy can possibly be probed by future 
cosmological/astrophysical observations 
(see, e.g., \cite{Hann06}).

\section{Conclusion}

\hskip 1.0cm The knowledge of neutrino mass spectrum is 
decisive for the understanding of the origin 
of neutrino masses and mixing. From the data
of neutrino oscillation experiments the two 
independent neutrino mass-squared differences 
in the case of 3-neutrino mixing, 
which are responsible for the solar and atmospheric 
neutrino oscillations, were determined. 
However, the existing data does not 
allow to determine the sign of the neutrino mass squared
difference driving the atmospheric neutrino oscillations,
${\rm sgn}(\Delta m_{23}^{2})$.
Thus, we can have $\Delta m_{23}^{2} > 0$, corresponding
(in the standardly used convention) to
{\it neutrino mass spectrum with normal ordering 
of neutrino masses}, $m_1 < m_2 < m_3$;
or $\Delta m_{23}^{2} < 0$, which is 
associated with {\it neutrino mass spectrum 
with inverted ordering of neutrino masses},
$m_3 < m_1 < m_2$.
In both cases the minimal value of the 
heaviest neutrino mass is given by 
$m^{\rm{min}}_{3(2)}\simeq 
\sqrt{|\Delta m^{2}_{23}|}\sim 5\cdot 10^{-2}$ eV.
We do not know at present the mass of the 
lightest neutrino ${\rm min}(m_j)$ as well.
The existing data from $\beta$-decay experiments and 
the cosmological data allow us to obtain only 
an upper bound on ${\rm min}(m_j)$.

  Depending on the value of the mass of the lightest neutrino, 
there are three possible characteristic
types of neutrino mass spectrum: 
{\it quasi-degenerate}, in which 
the lightest neutrino mass is relatively large, 
$m^2_{1(3)} \gg |\Delta m^{2}_{23}|$;
and two spectra in which the lightest 
neutrino mass is negligibly small: 
{\it normal hierarchical} with $\Delta m^{2}_{23} > 0$
and $m_{1}\ll\sqrt{\Delta m^{2}_{12}}$,
and {\it inverted hierarchical}, with 
$\Delta m^{2}_{23} < 0$ and
$m_{3}\ll \sqrt{|\Delta m^{2}_{23}|}$.
The future KATRIN 
$\beta$- decay experiment will probe 
the quasi-degenerate neutrino mass spectrum.
If  effect of neutrino mass will be 
observed in this experiment, the lightest 
neutrino mass will be measured. However, the 
${\rm sgn}(\Delta m_{23}^{2})$, and thus
the character of neutrino mass spectrum -
with normal or inverted ordering of neutrino masses, 
will not be determined in this case.

  If in the KATRIN experiment the effect of nonzero 
neutrino mass will not be observed, 
it will be crucial to improve the 
sensitivity of the $\beta$-decay experiments 
by approximately a factor of four,
which would permit to probe values
$m^{\rm{min}}_{3(2)}\cong \sqrt{|\Delta m^{2}_{23}|}$.
As we have shown, the non-observation 
of the effect of neutrino mass in a
$\beta$-decay experiment having a sensitivity to 
$\sqrt{|\Delta m^{2}_{23}|}$ 
would imply that the neutrino mass spectrum is
of the {\it normal hierarchical} type, i.e., that 
$\Delta m^{2}_{23} > 0$ and $m_1 \ll m_2 \ll m_3$,
independently of whether the massive 
neutrinos are Dirac or Majorana particles.
If the spectrum of neutrino masses is of the 
{\it inverted hierarchical} type,
the effect of neutrino mass must be observed 
in such an experiment.

  It follows from our analysis, however, that 
the interpretation of a positive result of 
a $\beta$-decay experiment with a sensitivity to
a neutrino mass $\sim \sqrt{|\Delta m^{2}_{23}|}$
will not be unique in what regards the 
${\rm sgn}(\Delta m^{2}_{23})$
and the type of neutrino mass spectrum.
The distortion of the electron spectrum
observed in such an experiment 
could be due either to i) spectrum 
with {\it inverted neutrino mass ordering}
($\Delta m^{2}_{23} < 0$) of two possible types,
a) {\it inverted hierarchical}, $m_3 \ll m_1 < m_2$, or
b) {\it with partial inverted hierarchy}, $m_3 < m_1 < m_2$;
or to ii) spectrum with {\it normal neutrino mass ordering}
($\Delta m^{2}_{23} > 0$), 
but with {\it partial neutrino mass hierarchy},
$m_1 < m_2 < m_3$.

\vspace{0.3cm}
{\bf Acknowledgements.} M.M. would like to thank 
the Elementary Particle Physics Sector of SISSA, 
where this work was done, for kind hospitality.
S.M.B. acknowledges the support of the Italian MIUR 
program ``Rientro dei Cervelli''. 
This work was also supported in part by the Italian MIUR 
program ``Fisica Astroparticellare'' (M.M. and S.T.P.),
by the Bulgarian National Science Fund
under the contract Ph-09-05 (M.M.) and by the 
the European Network of Theoretical Astroparticle Physics 
ILIAS/N6 under the contract RII3-CT-2004-506222
(S.M.B. and S.T.P.).

\end{document}